\documentclass[preprint,preprintnumbers, prd, floatfix, superscriptaddress,nofootinbib] {revtex4-1}
\usepackage{epsfig}
\usepackage{subfigure}
\usepackage{dcolumn}
\usepackage{bm}
\usepackage[usenames ,dvipsnames]{xcolor}
\usepackage{slashed}
\usepackage{graphicx,color}
\begin{document}
\title{Two-body charmed baryon decays involving decuplet baryon\\
in the quark-diagram scheme}

\author{Y.K. Hsiao}
\email{yukuohsiao@gmail.com}
\affiliation{School of Physics and Information Engineering, Shanxi Normal University, Linfen 041004, China}

\author{Qian Yi}
\affiliation{School of Physics and Information Engineering, Shanxi Normal University, Linfen 041004, China}

\author{Shu-Ting Cai}
\affiliation{School of Physics and Information Engineering, Shanxi Normal University, Linfen 041004, China}

\author{H.J. Zhao}
\affiliation{School of Physics and Information Engineering, Shanxi Normal University, Linfen 041004, China}

\date{\today}

\begin{abstract}
In the quark-diagram scheme,
we study the charmed baryon decays of ${\bf B}_c\to {\bf B}^* M$,
where ${\bf B}_c$ is $\Lambda_c^+$ or $\Xi_c^{+(0)}$,
together with ${\bf B}^*$ ($M$) the decuplet baryon (pseudoscalar meson).
It is found that only two $W$-exchange processes are allowed to contribute to ${\bf B}_c\to {\bf B}^* M$.
Particularly,
we predict ${\cal B}(\Lambda_c^+ \to \Sigma^{*0(+)} \pi^{+(0)})=(2.8\pm 0.4)\times 10^{-3}$,
which respects the isospin symmetry.
Besides, we take into account the $SU(3)$ flavor symmetry breaking,
in order to explain the observation of ${\cal B}(\Lambda_c^+\to \Sigma^{*+}\eta)$.
For the decays involving $\Delta^{++}(uuu)$,
we predict ${\cal B}(\Lambda_c^+\to \Delta^{++} \pi^-,\Xi_c^+ \to \Delta^{++} K^-)
=(7.0\pm 1.4,13.5\pm 2.7)\times 10^{-4}$ as the largest branching fractions in the singly
Cabibbo-suppressed $\Lambda_c^+,\Xi_c^+\to{\bf B}^*M$ decay channels, respectively,
which are accessible to the LHCb, BELLEII and BESIII experiments.
\end{abstract}

\maketitle
\section{introduction}
To determine the mass and lifetime of the $\Lambda_b$ baryon,
$\Lambda_c^+$ is often taken as the final state in the $\Lambda_b$ decays~\cite{pdg},
which involves the favored Cabibbo-Kobayashi-Maskawa  (CKM) matrix elements
for bigger branching fractions.
With the higher precision in the recent years~\cite{Zupanc:2013iki,Ablikim:2015flg},
the subsequent $\Lambda_c^+\to p K^-\pi^+$ decay
has helped to make more accurate observations for the $\Lambda_b$ decays,
which receives the significant contribution from
$\Lambda_c^+\to \Delta^{++}K^-,\Delta^{++}\to p \pi^+$.
Similarly, one uses $\Xi_b^0\to \Xi_c^+\pi^-$
to determine the $\Xi_b^0$ lifetime, whereas we find that
the subsequent process $\Xi_c^+\to pK^-\pi^+$ and
its resonant contribution from $\Xi_c^+\to \Delta^{++}K^-,\Delta^{++}\to p \pi^+$
have not been well studied yet~\cite{pdg}.
Therefore, ${\bf B}_c\to{\bf B}^* M$ plays the key role in
the precision measurements for the multi-body decays of beauty and charm baryons,
where ${\bf B}_c=(\Lambda_c^+,\Xi_c^{+(0)})$,
${\bf B}^*$ the decouplet baryon and $M$ the meson state,
such as $\Lambda_c^+(\Xi_c^+)\to \Delta^{++}K^-$.

The ${\bf B}_c\to{\bf B}^* M$ decays are not richly observed.
Therefore, it is still unclear how the ${\bf B}_c\to{\bf B}^* M,{\bf B}^*\to {\bf B}M'$ decays
mix with the non-resonant contributions to ${\bf B}_c\to {\bf B}MM'$.
In addition, ${\bf B}_c\to{\bf B}V,V\to MM'$ with $V$ the vector meson
causes more complicated mixtures~\cite{pdg}.
The $SU(3)$ flavor ($SU(3)_f$) symmetry has been widely applied to 
the charmed baryon decays~\cite{Savage:1989qr,Savage:1991wu,
He:2000ys,Fu:2003fy,Hsiao:2015iiu,Lu:2016ogy,Geng:2019awr,
Geng:2017esc,Geng:2018plk,Geng:2017mxn,Geng:2018upx,Hsiao:2019yur,Pan:2020qqo}.
By well explaining the data, 
the flavor symmetry does not appear to be severely broken
in ${\bf B}_c\to{\bf B}M$~\cite{Zhao:2018mov}, where $\bf B$ denotes the octet baryon.
By contrast, ${\cal B}(\Lambda_c^+\to\Sigma^{*+} \eta)$ not well interpreted
by the $SU(3)_f$ symmetry might hint the broken effect
in ${\bf B}_c\to {\bf B}^* M$~\cite{Geng:2019awr}, which could be 
as large as that in the $D$ meson decays~\cite{Cheng:2012xb,Li:2012cfa,
Li:2013xsa,Cheng:2019ggx}.

For a better understanding of the hadronization in ${\bf B}_c\to{\bf B}^* M$,
there have been some theoretical attempts, which are in terms of
the pole model, quark model and irreducible $SU(3)$ flavor ($SU(3)_f$) symmetry~\cite{Xu:1992sw,
Korner:1992wi,Sharma:1996sc,Geng:2017mxn,Savage:1989qr,Savage:1991wu,Geng:2019awr}.
Particularly,
the quark-diagram scheme with the topological $SU(3)_f$ symmetry
provides a clear picture for the decay processes~\cite{Kohara:1991ug,
Chau:1995gk,He:2018joe,Zhao:2018mov}.
Due to the fact that ${\bf B}^*$ is a spin-3/2 baryon with totally symmetric quark contents,
it can be shown that the topological diagrams involving the flavor anti-symmetric quark pair in ${\bf B}^*$
are forbidden or suppressed.
Therefore, we purpose to use the quark-diagram scheme
to relate all possible ${\bf B}_c\to{\bf B}^*M$ decay channels.
With the existing data, we will perform the numerical analysis, and
determine different topological contributions.
We can hence test the validity of the topological scheme, which involves
the $SU(3)_f$ symmetry and its broken effect.
Furthermore, we will give predictions for ${\cal B}({\bf B}_c\to {\bf B}^* M)$
to be compared to the future measurements, which can help to clarify how
${\bf B}_c\to{\bf B}^*M,{\bf B}^*\to {\bf B}M'$ mixes with
${\bf B}_c\to {\bf B}V,V\to MM'$ and the non-resonant configuration in ${\bf B}_c\to {\bf B}MM'$.

\section{Formalism}
\subsection{Effective Hamiltonian in the flavor symmetry}
To study the two-body charmed baryon decays,
the corresponding quark-level effective Hamiltonian is given by~\cite{Buras:1998raa}
\begin{eqnarray}\label{Heff}
{\cal H}_{eff}&=&\frac{G_F}{\sqrt 2}\sum_{i=1,2} c_i
\left(\lambda_a O_i^a+\lambda_p O_i^p+\lambda_c O_i^c\right)\,,
\end{eqnarray}
with $\lambda_{(a,p,c)}\equiv (V_{cs}^*V_{ud},V_{cp}^*V_{up},V_{cd}^*V_{us})$
and $p=(d,s)$,
where $G_F$ is the Fermi constant, and $c_i$ the Wilson coefficients.
The current-current operators $O_i^{(a,p,c)}$ are written as
\begin{eqnarray}\label{Oa}
&&
O_1^a=(\bar u d)(\bar s c)\,,\;
O_2^a=(\bar u_\beta d_\alpha)(\bar s_\alpha c_\beta)\,,\nonumber\\
&&
O_1^p=(\bar u p)(\bar p c)\,,\;
O_2^p=(\bar u_\beta p_\alpha)(\bar p_\alpha c_\beta)\,,\nonumber\\
&&
O_1^c=(\bar u s)(\bar d c)\,,\;
O_2^c=(\bar u_\beta s_\alpha)(\bar d_\alpha c_\beta)\,,
\end{eqnarray}
where $(\bar q_1 q_2)=\bar q_1\gamma_\mu(1-\gamma_5)q_2$,
and the subscripts $(\alpha,\beta)$ denote the color indices.
With $s_c\equiv \sin\theta_c\simeq 0.22$,
where $\theta_c$ denotes the Cabibbo angle for the quark-mixing in the weak interaction,
the decays with $|\lambda_{(a,p,c)}|\simeq (1,s_c,s_c^2)$
are regarded as the Cabibbo-allowed (CA), singly Cabibbo-suppressed (SCS) and
doubly Cabibbo-suppressed (DCS) processes, respectively.

For the lowest-lying anti-triplet charmed baryon states
$\Xi_c^0$, $\Xi_c^+$ and $\Lambda_c^+$ that consist of
$(ds-sd)c$, $(su-us)c$ and $(ud-du)c$, respectively,
we present them as
\begin{eqnarray}\label{b_octet}
{\bf B}_c&=&\left(\begin{array}{ccc}
0& \Lambda_c^+ & \Xi_c^+\\
-\Lambda_c^+&0&\Xi_c^0 \\
-\Xi_c^+&-\Xi_c^0&0
\end{array}\right)\,.
\end{eqnarray}
The pseudoscalar meson states are given by
\begin{eqnarray}\label{M_octet}
M=\left(\begin{array}{ccc}
\frac{1}{\sqrt{2}}(\pi^0+ c\phi\eta +s\phi\eta' ) & \pi^- & K^-\\
\pi^+ & \frac{-1}{\sqrt{2}}(\pi^0- c\phi\eta -s\phi\eta') & \bar K^0\\
K^+ & K^0& -s\phi\eta +c\phi\eta'
\end{array}\right)\,,
\end{eqnarray}
where $(\eta,\eta')$ mix with
$\eta_q=\sqrt{1/2}(u\bar u+d\bar d)$ and $\eta_s=s\bar s$.
The mixing angle $\phi=(39.3\pm1.0)^\circ$ in $(s\phi,c\phi)\equiv (\sin\phi,\cos\phi)$
comes from the mixing matrix,
given by~\cite{FKS}
\begin{eqnarray}\label{eta_mixing}
\left(\begin{array}{c} \eta \\ \eta^\prime \end{array}\right)
=
\left(\begin{array}{cc} \cos\phi & -\sin\phi \\ \sin\phi & \cos\phi \end{array}\right)
\left(\begin{array}{c} \eta_q \\ \eta_s \end{array}\right).
\end{eqnarray}
The decuplet baryons are written as
\begin{eqnarray}\label{B_10}
&&{\bf B}^*=\frac{1}{\sqrt{3}}
\left(\begin{array}{ccc}
\left(\begin{array}{ccc}
\sqrt{3}\Delta^{++}&\Delta^+ & \Sigma^{* +}\\
\Delta^+ &\Delta^0 & \frac{\Sigma^{* 0}}{\sqrt{2}}\\
\Sigma^{* +}& \frac{\Sigma^{* 0}}{\sqrt{2}}& \Xi^{* 0}
\end{array}\right),
\left(\begin{array}{ccc}
\Delta^+ &\Delta^0 & \frac{\Sigma^{* 0}}{\sqrt{2}}\\
\Delta^0 &\sqrt{3}\Delta^-& \Sigma^{* -}\\
\frac{\Sigma^{* 0}}{\sqrt{2}}&\Sigma^{* -}&\Xi^{* -}
\end{array}\right),
\left(\begin{array}{ccc}
\Sigma^{* +}& \frac{\Sigma^{* 0}}{\sqrt{2}}&\Xi^{* 0}\\
\frac{\Sigma^{* 0}}{\sqrt{2}}&\Sigma^{* -}&\Xi^{* -}\\
\Xi^{* 0}&\Xi^{* -}&\sqrt{3}\Omega^-
\end{array}\right)
\end{array}\right).
\end{eqnarray}
By neglecting the Lorentz indices, ${\cal H}_{eff}$
for the $c\to q_i\bar q_j q_k$ transition can be presented
with the tensor notation, $H^{ki}_j$, and
the nonzero entries are given by~\cite{He:2018joe}
\begin{eqnarray}\label{Hijk}
H^{31}_2=\lambda_a, H^{21}_2=\lambda_d,
H^{31}_3=\lambda_s, H^{21}_3=\lambda_c\,.
\end{eqnarray}

\subsection{The quark-diagram scheme}
%
\begin{figure}
\includegraphics[width=0.31\textwidth]{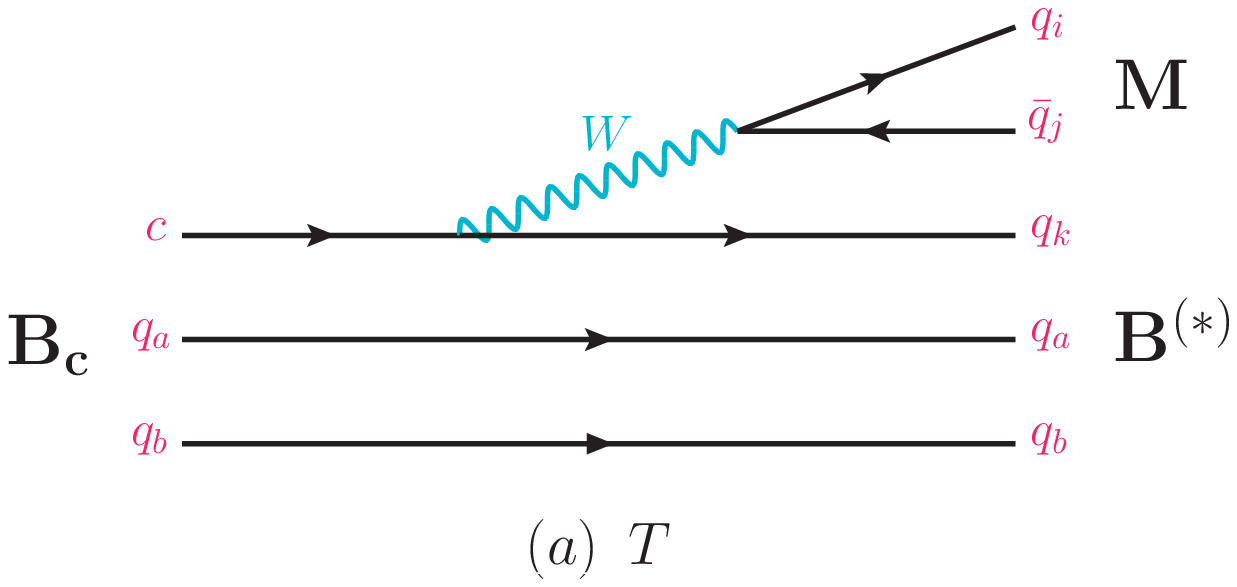}
\includegraphics[width=0.31\textwidth]{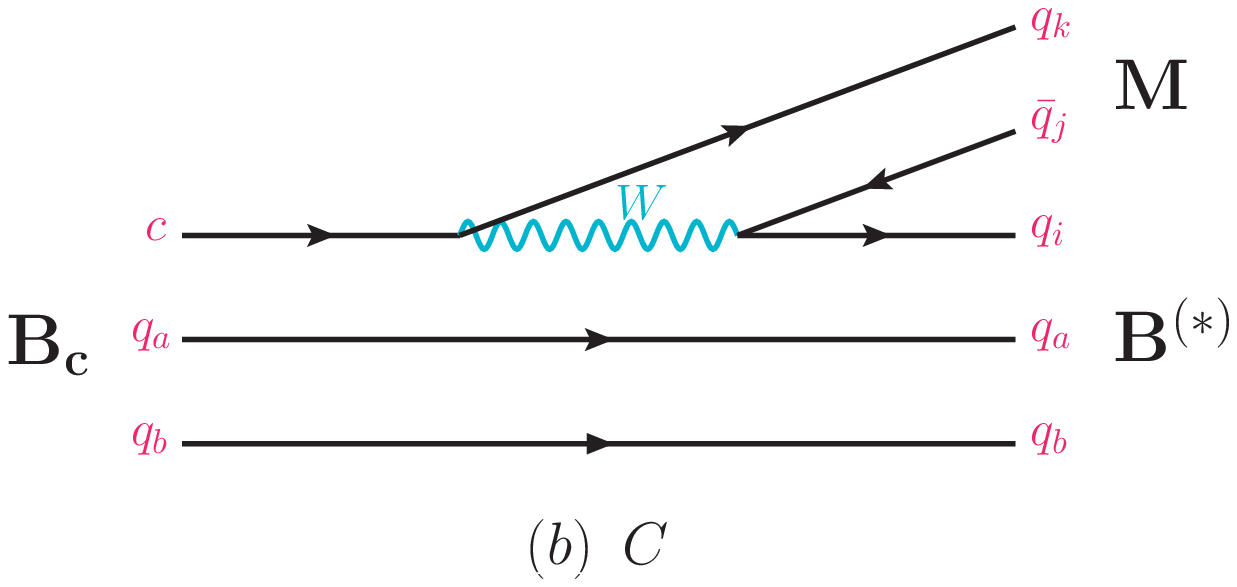}
\includegraphics[width=0.31\textwidth]{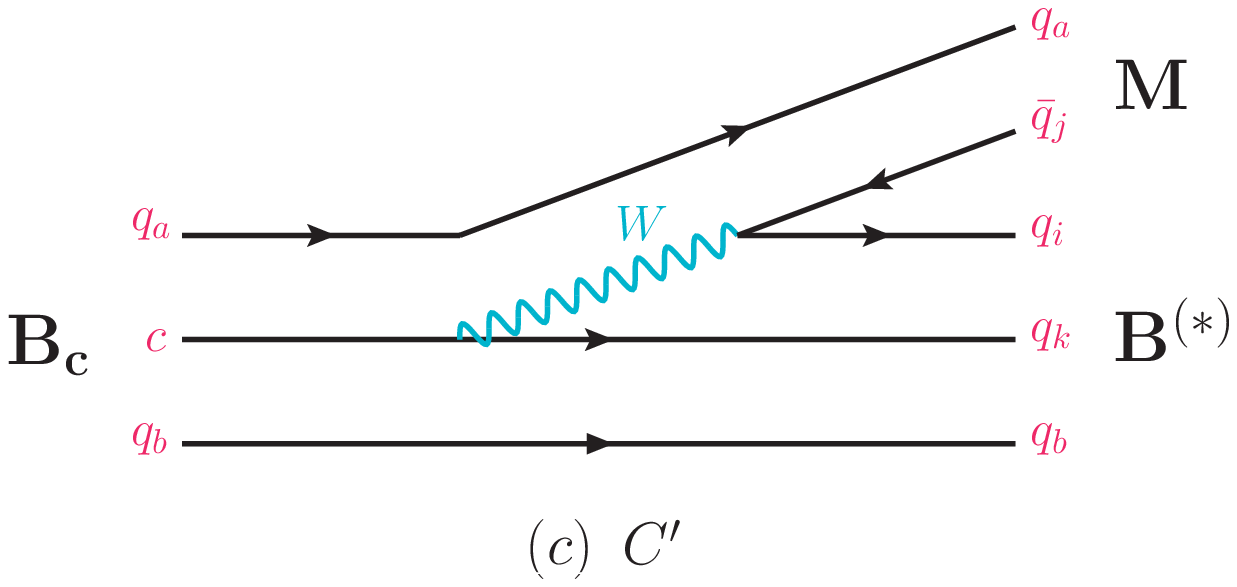}
\includegraphics[width=0.31\textwidth]{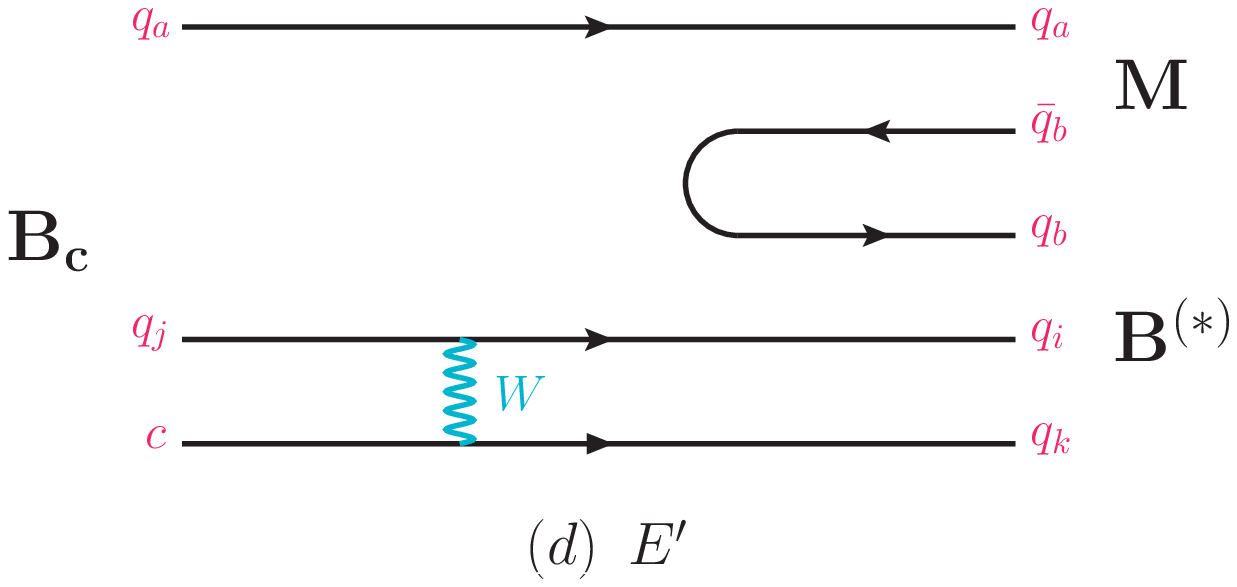}
\includegraphics[width=0.31\textwidth]{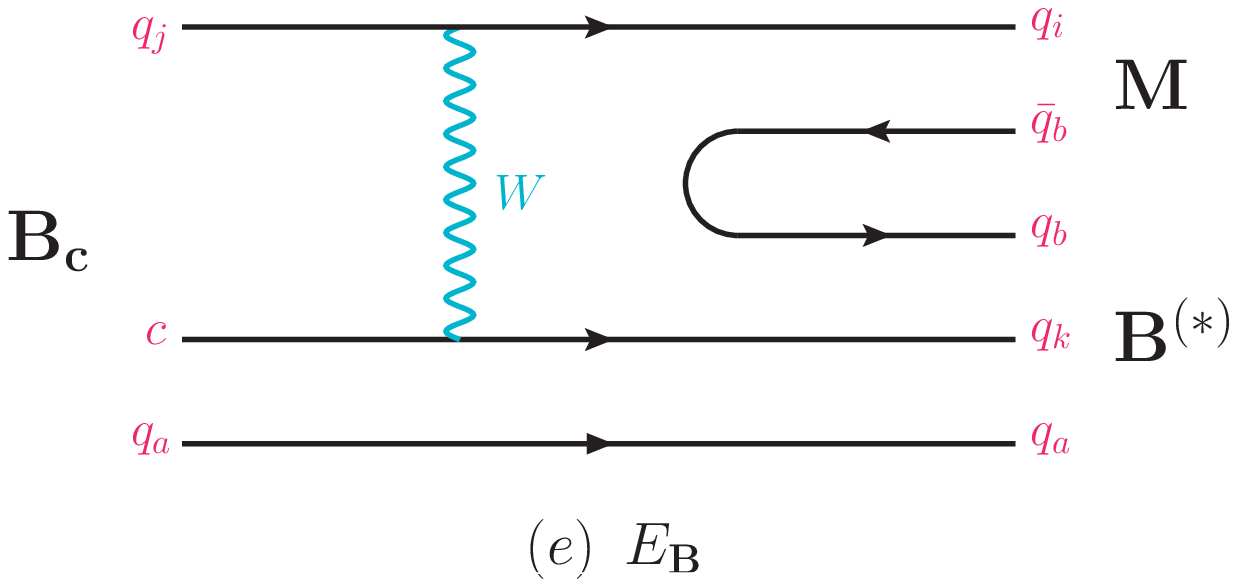}
\includegraphics[width=0.31\textwidth]{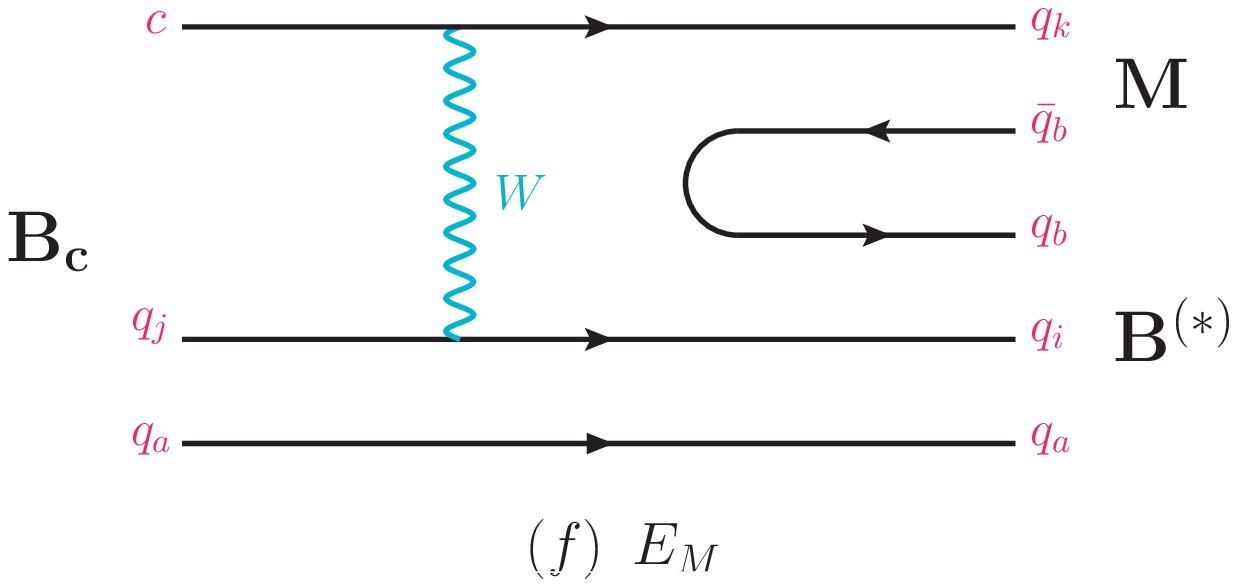}
\caption{Topological diagrams for the ${\bf B}_c\to {\bf B}^{(*)} M$ decays.}\label{figT}
\end{figure}
%
In the quark-diagram scheme,
there exist six different topological diagrams
for the ${\bf B}_c\to {\bf B}^{(*)}M$ decays,
as drawn in Figs.~\ref{figT}(a,b,c) and \ref{figT}(d,e,f),
parameterized as the topological amplitudes
$(T,C,C')$ and $(E',E_{\bf B},E_M)$, respectively~\cite{Zhao:2018mov}.
More explicitly,
$T$ and $C^{(\prime)}$ proceed with the $W$-boson emission ($W_{\text{EM}}$).
By exchanging the $W$ boson ($W_{\text{EX}}$), it gives rise to $E'$ and $E_{{\bf B}(M)}$.
Since only $(T,C)$ can be decomposed of two separate matrix elements
based on the factorization, that is, $(T,C)\propto
\langle M|\bar q_1 q_2|0\rangle \langle {\bf B}^{(*)}|\bar q_3 c|{\bf B}_c\rangle$~\cite{Hsiao:2020gtc},
one classifies $(T,C)$ and $(C',E',E_{M,{\bf B}})$
as the factorizable and non-factorizable amplitudes, respectively.

Furthermore, it is found in Figs.~\ref{figT}(a,b) that
${\bf B}_c$ with $(q_a q_b-q_b q_a)c$ cannot be turned into ${\bf B}^*(q_a q_b q_{k(i)})$,
where $q_a q_b q_{k(i)}$ are totally symmetric,
such that $(T,C)$ give no contributions to ${\bf B}_c\to {\bf B}^* M$.
Thus, the ${\bf B}_c\to{\bf B}^*M$ decays are purely non-factorizable processes.
In addition, $C'$ and $E'$ are suppressed in ${\bf B}_c\to {\bf B}^* M$~\cite{Kohara:1991ug},
which is in accordance with the K$\ddot{\text{o}}$rner-Pati-Woo theorem~\cite{KPW}.
With the current-current structure of $(\bar q_i q_j)_{V-A}(\bar q_k c)_{V-A}$
in Eq.~(\ref{Oa}), $q_i$ and $q_k$ are color anti-symmetric.
When combined as the constituents of the baryon,
$q_{i,k}$ are flavor anti-symmetric, such that
the topological diagrams $(C',E')$ in Figs.~\ref{figT}(c,d)
contribute to ${\bf B}_c\to{\bf B}M$, instead of ${\bf B}_c\to{\bf B}^* M$.
Consequently, we are left with the $W_{\text{EX}}$ topological diagrams
$(E_{\bf B},E_M)$ in Figs.~\ref{figT}(e,f) for ${\bf B}_c\to{\bf B}^* M$.

To proceed, we derive the amplitudes as
${\cal A}({\bf B}_c\to {\bf B}^*M)=({G_F}/{\sqrt 2}) T({\bf B}_c\to {\bf B}^*M)$.
Explicitly, the $T$ amplitudes ($T$-amps)
read~\cite{Kohara:1991ug,He:2018joe,Zhao:2018mov}
\begin{eqnarray}\label{amp1}
T({\bf B}_c\to {\bf B}^*M)=
E_{\bf B}^{(s)} ({{\bf B}_c})^{ja} H^{ki}_j({\bf B}^*)_{kab}(M)_i^b
+E_M^{(s)} ({{\bf B}_c})^{ja} H^{ki}_j({\bf B}^*)_{iab}(M)_k^b\,,
\end{eqnarray}
where the parameters $E_{{\bf B},M}^{(s)}$ correspond to the topological diagrams
in Figs.~\ref{figT}e and \ref{figT}f, respectively.
The $W_{\text{EX}}$ decay process needs an additional quark pair
from $g\to q\bar q$, where $q\bar q$ could be $u\bar u$, $d\bar d$ or $s\bar s$.
To take into account the broken $SU(3)_f$ symmetry, 
$E_{{\bf B}(M)}$ with $g\to s\bar s$ can be more specifically denoted by $E_{{\bf B}(M)}^s$.
Under the exact $SU(3)_f$ symmetry, it leads to
$E_{{\bf B}(M)}^s=E_{{\bf B}(M)}$~\cite{Chau:1995gk}.
In Tables~\ref{tab1}, \ref{tab2} and \ref{tab3}, we present
the full expansions of $T({\bf B}_c\to {\bf B}^*M)$
for the CA, SCS and DCS decay modes, respectively.
For the branching fractions,
we use the equation for the two-body decays,
given by~\cite{pdg}
\begin{eqnarray}\label{p_space}
&&{\cal B}({\bf B}_c\to{\bf B}^* M)=
\frac{G_F^2|\vec{p}_{{\bf B}^*}|\tau_{{\bf B}_c}}{16\pi m_{{\bf B}_c}^2 }
|T({\bf B}_c\to{\bf B}^* M)|^2\,,\nonumber\\
&&|\vec{p}_{{\bf B}^*}|=\frac{
\sqrt{(m_{{\bf B}_c}^2-m_+^2)(m_{{\bf B}_c}^2-m_-^2)}}{2 m_{{\bf B}_c}}\,,
\end{eqnarray}
with $m_\pm=m_{{\bf B}^*}\pm m_M$,
where $\tau_{{\bf B}_c}$ stands for the ${\bf B}_c$ baryon lifetime.

\section{Numerical Analysis and Discussions}
\begin{table}[b!]
\caption{Cabibbo-allowed ${\bf B}_c\to {\bf B}^* M$ decays.}\label{tab1}
\scriptsize
\begin{tabular}{|l|l|c|c|c|}
\hline
Decay modes& $T$-amp
&$10^{3}{\cal B}~(S1,\,S2)$~[our work]
&$10^{3}{\cal B}~(S_{pm},\,S_{em})$~\cite{Geng:2019awr}
&$10^{3}{\cal B}_{ex}$~\cite{pdg,Li:2019atu}\\
\hline\hline
$\Lambda_c^+ \to \Delta^{++} K^-$
&$-\lambda_aE_M$
&$(12.0\pm 2.2,\,11.7\pm 2.3)$
&$(15.3\pm 2.4,\,12.4\pm 1.0)$
&$10.8\pm 2.5$\\
$\Lambda_c^+ \to \Delta^+ \bar K^0$
&$-\lambda_a\frac{1}{\sqrt 3}E_M$
&$(4.0\pm 0.7,\,3.9\pm 0.8)$
&$(5.1\pm 0.8,\,4.1\pm 0.3)$
&----- \\
$\Lambda_c^+ \to \Sigma^{*0} \pi^+$
&$-\lambda_a\frac{1}{\sqrt 6}E_{\bf B}$
&$(2.9\pm 0.4,\,2.8\pm 0.4)$
&$(2.2\pm 0.4,\,2.1\pm 0.2)$
&----- \\
$\Lambda_c^+ \to \Sigma^{*+} \pi^0$
&$-\lambda_a\frac{1}{\sqrt 6}E_{\bf B}$
&$(2.9\pm 0.4,\,2.8\pm 0.4)$
&$(2.2\pm 0.4,\,2.1\pm 0.2)$
&----- \\
$\Lambda_c^+ \to \Sigma^{*+} \eta$
&$-\lambda_a\frac{1}{\sqrt 6}(E_{\bf B}c\phi-\sqrt 2 E_M^{(s)} s\phi)$
&$(5.3\pm 0.8,\,7.3\pm 1.5)$
&$(3.1\pm 0.6,\,6.2\pm 0.5)$
&$9.1\pm2.0$\\
$\Lambda_c^+ \to \Sigma^{*+} \eta^\prime $
&$-\lambda_a\frac{1}{\sqrt 6}(E_{\bf B} s\phi+\sqrt 2 E_M^{(s)} c\phi)$
&(0,\,0)
&-----
&----- \\
$\Lambda_c^+ \to  \Xi^{*0} K^+$
&$-\lambda_a\frac{1}{\sqrt 3}E_{\bf B}$
&$(3.9\pm 0.6,\,3.9\pm 0.6)$
&$(1.0\pm 0.2,\,4.1\pm 0.3)$
&$4.3\pm 0.9$\\
\hline
$\Xi_c^0 \to \Sigma^{*+} K^-$
&$\lambda_a\frac{1}{\sqrt 3}E_M$
&$(1.8\pm 0.3,\,1.7\pm 0.3)$
&$(3.1\pm 0.5,\,2.3\pm 0.2)$
&----- \\
$\Xi_c^0 \to \Sigma^{*0} \bar{K}^0$
&$\lambda_a\frac{1}{\sqrt 6}E_M$
&$(0.9\pm 0.2,\,0.9\pm 0.2)$
&$(1.6\pm 0.2,\,1.2\pm 0.1)$
&----- \\
$\Xi_c^0 \to \Xi^{*-}\pi^+$
&$\lambda_a\frac{1}{\sqrt 3}E_{\bf B}$
&$(2.6\pm 0.4,\,2.5\pm 0.4)$
&$(2.8\pm 0.5,\,2.3\pm 0.2)$
&----- \\
$\Xi_c^0 \to \Xi^{*0} \pi^0$
&$\lambda_a\frac{1}{\sqrt 6}E_{\bf B}$
&$(1.3\pm 0.2,\,1.3\pm 0.2)$
&$(1.4\pm 0.2,\,1.2\pm 0.1)$
&----- \\
$\Xi_c^0 \to \Xi^{*0} \eta $
&$\lambda_a\frac{1}{\sqrt 6}(E_{\bf B}c\phi-\sqrt2 E_M^{(s)} s\phi)$
&$(2.4\pm 0.4,\,3.4\pm 0.7)$
&$(2.1\pm 0.4,\,3.5\pm 0.3)$
&----- \\
$\Xi_c^0 \to \Xi^{*0} \eta^\prime $
&$\lambda_a\frac{1}{\sqrt 6}(E_{\bf B} s\phi+\sqrt2 E_M^{(s)} c\phi)$
&$(0.01\pm 0.04,\,0.08\pm 0.10)$
&-----
&----- \\
$\Xi_c^0 \to \Omega^- K^+$
&$\lambda_aE_{\bf B}^{(s)}$
&$(4.8\pm 0.7,\,4.8\pm 0.7)$
&$(2.3\pm 0.5,\,7.0\pm 0.6)$
&$4.2\pm 1.0$\\
\hline
$\Xi_c^+ \to \Sigma^{*+} \bar{K}^0$
&0
&0
&0
&$28.6\pm 16.8$\\
$\Xi_c^+ \to \Xi^{*0} \pi^+$
&0
&0
&0
&$<4.0$\\
\hline
\end{tabular}
\end{table}
In the numerical analysis,
we adopt the Cabibbo-Kobayashi-Maskawa (CKM) matrix elements as~\cite{pdg}
\begin{eqnarray}\label{B1}
&&(V_{cs},V_{ud},V_{us},V_{cd})=(1-\lambda^2/2,1-\lambda^2/2,\lambda,-\lambda)\,,
\end{eqnarray}
with $\lambda=s_c=0.22453\pm 0.00044$
in the Wolfenstein parameterization. Besides,
the ${\bf B}_c$ and ${\bf B}^*$ masses, together with
the lifetime for ${\bf B}_c$, are adopted from the PDG~\cite{pdg}.
We perform a minimum $\chi^2$-fit
with $\chi^2=\sum({\cal B}_{th}-{\cal B}_{ex})^2/\sigma_{ex}^{2}$,
where ${\cal B}_{th(ex)}$ represents the
theoretical (experimental) input of the branching ratio,
and $\sigma_{ex}$ the experimental error.
We calculate ${\cal B}_{th}$ with the equation in Eq.~(\ref{p_space}),
together with $({\cal B}_{ex},\sigma_{ex})$ from Table~\ref{tab1}.
Note that ${\cal B}(\Xi_c^+ \to \Sigma^{*+} \bar K^0,\Xi^{*0} \pi^+)$
are not involved in the fit.

We use two scenarios for the global fit.
In the first scenario~($S1$), we take $E_{{\bf B}(M)}^s=E_{{\bf B}(M)}$
under the exact $SU(3)_f$ symmetry.
Since $E_{\bf B}$ and $E_{M}$ are complex numbers,
it leads to three independent parameters, given by
\begin{eqnarray}
|E_{\bf B}|,\,|E_{M}| e^{i\delta_{E_M}}\,,
\end{eqnarray}
where $E_{\bf B}$ is set to be real, and $\delta_{E_{M}}$ is a relative strong phase.
Using the $\chi^2$-fit, we extract that
\begin{eqnarray}\label{topo_fit}
&&(|E_{\bf B}|,|E_{M}|)=(0.41\pm 0.03,0.34\pm 0.03)\,\text{GeV}^3\,,\nonumber\\
&&\delta_{E_M}=(180.0\pm 35.8)^\circ\,,
\end{eqnarray}
with $\chi^2/n.d.f=4.5$, where $n.d.f=1$ is the number of the degrees of freedom.
For $\delta_{E_M}$, its information is from ${\cal B}(\Lambda_c^+ \to \Sigma^{*+} \eta)$.
Although $\delta_{E_M}=180^\circ$ has induced the largest positive interference
between $E_{\bf B}$ and $E_M$,
our result of ${\cal B}(\Lambda_c^+ \to \Sigma^{*+} \eta)=(5.3\pm 0.7)\times 10^{-3}$
is still shown to be in tension with the observation of $(9.1\pm2.0)\times 10^{-3}$.
Sizeably, it adds 3.6 to the total $\chi^2$ value.

Since $\Lambda_c^+ \to \Sigma^{*+} \eta$ is in association with $|E_M^s|$,
the tension hints the broken $SU(3)_f$ symmetry,
where $|E_M^s|$ is not equal to $|E_M|$.
On the other hand, ${\cal B}(\Xi_c^0 \to \Omega^- K^+)$
is fitted to agree with the data,
indicating that $|E_{\bf B}^s|$ is not deviating from $|E_{\bf B}|$.
Currently, the data points are not sufficient for an independent extraction of $|E_M^s|$.
We hence adopt the numerical results from the two-body $D$ meson decays, where
the similar $W_{\text{EX}}$ contributions have been found to induce
the severe $SU(3)_f$ symmetry breaking~\cite{Cheng:2012xb,Li:2012cfa,Li:2013xsa,Cheng:2019ggx}.
In the second scenario~($S2$),
we take $|E_M^s|=n_q\times |E_M|$ and $|E_{\bf B}^s|\simeq |E_{\bf B}|$,
with $n_q=1.4$ adopted from~\cite{Cheng:2019ggx}.
Consequently, we obtain
\begin{eqnarray}\label{topo_fit2}
&&(|E_{\bf B}|,|E_{M}|)=(0.40\pm 0.03,0.34\pm 0.03)\,\text{GeV}^3\,,\nonumber\\
&&\delta_{E_M}=(180.0\pm 46.8)^\circ\,,
\end{eqnarray}
where $\chi^2/n.d.f$ is reduced as 1.3.
As the demonstration, we obtain
${\cal B}(\Lambda_c^+ \to \Sigma^{*+} \eta)=(7.3\pm 1.5)\times 10^{-3}$,
which alleviates the deviation from the observation.
With the fit values of $(|E_{\bf B}|,|E_{M}|,\delta_{E_M})$ in $S1$ and $S2$,
we present the branching ratios of the ${\bf B}_c\to{\bf B}^* M$ decays
in Tables~\ref{tab1}, \ref{tab2} and \ref{tab3},
along with the recent theoretical results for comparison.
%
\begin{table}[b!]
\caption{Singly Cabibbo-suppressed ${\bf B}_c\to {\bf B}^* M$ decays.}\label{tab2}
\scriptsize

\begin{tabular}{|l|l|c|c|}
\hline
Decay modes& $T$-amp
&$10^{4}{\cal B}~(S1,\,S2)$~[our work]
&$10^{4}{\cal B}~(S_{pm},\,S_{em})$~\cite{Geng:2019awr}\\
\hline\hline
$\Lambda_c^+ \to \Delta^{++} \pi^-$
&$-\lambda_dE_M$
&$(7.2\pm 1.3,\,7.0\pm 1.4)$
&$(12.5\pm 2.0,\,6.6\pm 0.6)$\\
$\Lambda_c^+ \to \Delta^+ \pi^0$
&$-\lambda_d\frac{1}{\sqrt 6}( E_{\bf B}- E_M)$
&$(5.8\pm 0.9,\,5.6\pm 1.1)$
&$(8.3\pm 1.3,\,4.4\pm 0.4)$\\
$\Lambda_c^+ \to \Delta^0 \pi^+$
&$-\lambda_d\frac{1}{\sqrt 3}E_{\bf B}$
&$(3.5\pm 0.5,\,3.3\pm 0.5)$
&$(4.2\pm 0.7,\,2.2\pm 0.2)$\\
$\Lambda_c^+ \to \Delta^+ \eta$
&$-\lambda_d\frac{1}{\sqrt 6}(E_{\bf B}+E_M)c\phi$
&$(0.03\pm 0.28,\,0.02\pm 0.45)$
&-----\\
$\Lambda_c^+ \to \Delta^+ \eta^\prime$
&$-\lambda_d\frac{1}{\sqrt 6}(E_{\bf B}+E_M)s\phi$
&$(0.01\pm 0.09,\,0.01\pm 0.15)$
&-----\\
$\Lambda_c^+ \to \Sigma^{*+}  K^0$
&$-\lambda_d\frac{1}{\sqrt 3}E_M^{(s)}$
&$(1.8\pm 0.4,\,3.5\pm 0.7)$
&$(1.3\pm 0.2,\,2.2\pm 0.2)$\\
$\Lambda_c^+ \to \Sigma^{*0}  K^+$
&$-\lambda_d\frac{1}{\sqrt 6}E_{\bf B}^{(s)}$
&$(1.3\pm 0.2,\,1.3\pm 0.2)$
&$(0.7\pm 0.1,\,1.1\pm 0.1)$\\
\hline
$\Xi_c^0 \to \Delta^+ K^-$
&$-\lambda_s \frac{1}{\sqrt 3}E_M$
&$(1.1\pm 0.2,\,1.0\pm 0.2)$
&$(3.0\pm 0.5,\,1.2\pm 0.1)$\\
$\Xi_c^0 \to \Delta^0 \bar K^0$
&$-\lambda_s \frac{1}{\sqrt 3}E_M$
&$(1.1\pm 0.2,\,1.0\pm 0.2)$
&$(3.0\pm 0.5,\,1.2\pm 0.1)$\\
$\Xi_c^0 \to \Sigma^{*-} \pi^+$
&$\frac{1}{\sqrt 3}(\lambda_dE_{\bf B} -\lambda_sE_{\bf B})$
&$(6.1\pm 0.9,\,5.8\pm 0.9)$
&$(9.9\pm 1.6,\,4.9\pm 0.4)$\\
$\Xi_c^0 \to\Sigma^{*+} \pi^-$
&$\lambda_d \frac{1}{\sqrt 3}E_M$
&$(1.0\pm 0.2,\,1.0\pm 0.2)$ 
&$(2.5\pm 0.4,\,1.2\pm 0.1)$\\
$\Xi_c^0 \to \Sigma^{*0}\pi^0$
&$\frac{1}{\sqrt {12}}[\lambda_d(E_{\bf B}-E_M)-\lambda_sE_{\bf B}]$
&$(3.1\pm 0.4,\,2.9\pm 0.5)$
&$(5.6\pm 0.9,\,2.8\pm 0.2)$\\
$\Xi_c^0 \to \Sigma^{*0}\eta $
&$\frac{1}{\sqrt {12}}[\lambda_d(E_{\bf B}+E_M)c\phi+\lambda_s({\sqrt 2}E_M^{(s)}s\phi-E_{\bf B} c\phi)]$
&$(0.9\pm 0.2,\,1.2\pm 0.3)$
&$(1.1\pm 0.2,\,0.9\pm 0.1)$\\
$\Xi_c^0 \to \Sigma^{*0}\eta^\prime $
&$\frac{1}{\sqrt {12}}[\lambda_d(E_{\bf B}+E_M)s\phi-\lambda_s({\sqrt 2}E_M^{(s)}c\phi+E_{\bf B} s\phi)]$
&$(0.004\pm 0.120,\,0.050\pm 0.250)$
&-----\\
$\Xi_c^0 \to \Xi^{*0} K^0 $
&$\frac{1}{\sqrt 3}\lambda_dE_M^{(s)} $
&$(0.8\pm 0.2,\,1.6\pm 0.4)$
&$(0.9\pm 0.2,\,1.2\pm 0.1)$\\
$\Xi_c^0 \to \Xi^{*-} K^+$
&$\frac{1}{\sqrt 3}(\lambda_dE_{\bf B}^{(s)}-\lambda_sE_{\bf B}^{(s)})$
&$(4.6\pm 0.7,\,4.6\pm 0.7)$
&$(3.6\pm 0.6,\,4.9\pm 0.4)$\\
\hline
$\Xi_c^+ \to \Delta^{++} K^-$
&$-\lambda_sE_M$
&$(13.8\pm 2.5,\,13.5\pm 2.7)$
&$(35.0\pm 5.7,\,14.6\pm 1.2)$\\
$\Xi_c^+ \to \Delta^+ \bar K^0$
&$-\lambda_s\frac{1}{\sqrt 3}E_M$
&$(4.6\pm 0.8,\,4.5\pm 0.9)$
&$(11.7\pm 1.9,\,4.9\pm 0.4)$\\
$\Xi_c^+ \to \Sigma^{*+} \pi^0$
&$-\lambda_s\frac{1}{\sqrt 6}E_{\bf B}$
&$(3.4\pm 0.5,\,3.2\pm 0.5)$
&$(4.8\pm 0.8,\,2.4\pm 0.2)$\\
$\Xi_c^+ \to \Sigma^{*0} \pi^+$
&$-\lambda_s\frac{1}{\sqrt 6}E_{\bf B}$
&$(3.4\pm 0.5,\,3.2\pm 0.5)$
&$(4.8\pm 0.8,\,2.4\pm 0.2)$\\
$\Xi_c^+ \to \Sigma^{*+}\eta $
&$-\lambda_s\frac{1}{\sqrt 6}(E_{\bf B}c\phi-{\sqrt 2}E_M^{(s)} s\phi) $
&$(6.4\pm 1.0,\,9.1\pm 1.8)$
&$(8.7\pm 1.4,\,7.3\pm 0.6)$\\
$\Xi_c^+ \to \Sigma^{*+}\eta^\prime $
&$-\lambda_s\frac{1}{\sqrt 6}(E_{\bf B} s\phi+{\sqrt 2}E_M^{(s)} c\phi) $
&$(0.1\pm 0.3,\,0.6\pm 0.8)$
&-----\\
$\Xi_c^+ \to \Xi^{*0}  K^+$
&$-\lambda_s\frac{1}{\sqrt 3}E_{\bf B}^{(s)}$
&$(5.0\pm 0.8,\,5.0\pm 0.8)$
&$(3.5\pm 0.6,\,4.9\pm 0.4)$\\
\hline
\end{tabular}
\end{table}
%

We get some useful relations in the quark-diagram scheme.
For example,
we find out three triangle sum rules for ${\bf B}_c\to\Delta\pi$,
given by
\begin{eqnarray}
&&T(\Lambda_c^+\to \Delta^0\pi^+)-T(\Lambda_c^+\to\Delta^{++}\pi^-)
-\sqrt 6 T(\Lambda_c^+\to\Delta^+\pi^0)=0\,,
\nonumber\\
&&T(\Xi_c^+\to \Delta^0\pi^+)-T(\Xi_c^+\to\Delta^{++}\pi^-)
-\sqrt 6 T(\Xi_c^+\to\Delta^+\pi^0)=0\,,
\nonumber\\
&&T(\Xi_c^0\to \Delta^+\pi^-)-T(\Xi_c^0\to\Delta^{-}\pi^+)
-\sqrt 6 T(\Xi_c^0\to\Delta^0\pi^0)=0\,.
\end{eqnarray}
Besides, we obtain
\begin{eqnarray}\label{zeromode}
&&T(\Lambda_c^+ \to \Delta^+ K^0,\Delta^0  K^+)=0\,,\nonumber\\
&&T(\Xi_c^+ \to \Sigma^{*+} \bar{K}^0,\Xi^{*0} \pi^+)=0\,,
\end{eqnarray}
as the consequence of $C'$
being set to give no contribution to ${\bf B}_c\to {\bf B}^* M$.
Indeed, $C'$ is the only topology that takes part in the decays in Eq.~(\ref{zeromode}),
but suppressed due to the K$\ddot{\text{o}}$rner-Pati-Woo theorem~\cite{KPW}.
According to the other theoretical calculations~\cite{Xu:1992sw,Korner:1992wi,
Sharma:1996sc,Geng:2017mxn,Geng:2019awr},
${\cal B}(\Xi_c^+ \to \Sigma^{*+} \bar K^0,\Xi^{*0} \pi^+)=0$ is also predicted,
which supports that $C'=0$. Experimentally,
${\cal B}_{ex}(\Xi_c^+ \to \Sigma^{*+} \bar K^0,\Xi^{*0} \pi^+)$ in Table~\ref{tab2}
can be used to test the suppression.
With ${\cal B}(\Xi_c^+ \to \Xi^{*0} \pi^+)/{\cal B}(\Xi_c^+\to\Xi^-\pi^+\pi^+)<0.1$
and ${\cal B}(\Xi_c^+\to\Xi^-\pi^+\pi^+) =(2.86\pm 1.21\pm 0.38)\times 10^{-2}$~\cite{pdg,Li:2019atu},
we determine ${\cal B}_{ex}(\Xi_c^+ \to \Xi^{*0} \pi^+)<4.0\times 10^{-3}$,
which can be seen as the non-observation to agree with $T(\Xi_c^+ \to \Xi^{*0} \pi^+)=0$.
However, ${\cal B}_{ex}(\Xi_c^+ \to \Sigma^{*+} \bar K^0)=(2.9\pm 1.7)\times 10^{-2}$
seems to disagree with the prediction of ${\cal B}(\Xi_c^+ \to \Sigma^{*+} \bar K^0)=0$,
despite of the large uncertainty.
While the K$\ddot{\text{o}}$rner-Pati-Woo theorem
is regarded as a tree-level approximation, allowing possible corrections to $C'$ and $E'$,
we need more accurate observations to test if $C'(E')=0$.
%
\begin{table}[b!]
\caption{Doubly Cabibbo-suppressed ${\bf B}_c\to {\bf B}^* M$ decays.}\label{tab3}
\scriptsize
\begin{tabular}{|l|l|c|c|}
\hline
Decay modes& $T$-amp
&$10^{5}{\cal B}~(S1,\,S2)$~[our work]
&$10^{5}{\cal B}~(S_{pm},\,S_{em})$~\cite{Geng:2019awr}\\
\hline\hline
$\Lambda_c^+ \to \Delta^+ K^0$
&0
&0
&0\\
$\Lambda_c^+ \to \Delta^0  K^+$
&0
&0
&0\\
\hline
$\Xi_c^0 \to \Delta^+ \pi^-$
&$-\lambda_c\frac{1}{\sqrt 3}E_M$
&$(0.6\pm 0.1,\,0.6\pm 0.1)$
&$(2.2\pm 0.4,\,0.7\pm 0.1)$\\
$\Xi_c^0 \to \Delta^0 \pi^0$
&$-\lambda_c\frac{1}{\sqrt 6}(E_{\bf B}-E_M)$
&$(1.5\pm 0.2,\,1.4\pm 0.3)$
&$(4.3\pm 0.7,\,1.3\pm 0.1)$\\
$\Xi_c^0 \to \Delta^- \pi^+$
&$-\lambda_cE_{\bf B}$
&$(2.7\pm 0.4,\,2.5\pm 0.4)$
&$(6.5\pm 1.1,\,2.0\pm 0.2)$\\
$\Xi_c^0 \to \Delta^0 \eta$
&$-\lambda_c\frac{1}{\sqrt 6}(E_{\bf B}+E_M)c\phi$
&$(0.01\pm 0.07,\,0.01\pm 0.12)$
&-----\\
$\Xi_c^0 \to \Delta^0 \eta^\prime$
&$-\lambda_c\frac{1}{\sqrt 6}(E_{\bf B}+E_M)s\phi$
&$(0.003\pm 0.034,\,0.003\pm 0.055)$
&-----\\
$\Xi_c^0 \to \Sigma^{*0} K^0 $
&$-\lambda_c\frac{1}{\sqrt 6}E_M^{(s)}$
&$(0.2\pm 0.1,\,0.5\pm 0.1)$
&$(0.4\pm 0.1,\,0.3\pm 0.0)$\\
$\Xi_c^0 \to \Sigma^{*-} K^+ $
&$-\lambda_c\frac{1}{\sqrt 3}E_{\bf B}^{(s)}$
&$(0.7\pm 0.1,\,0.7\pm 0.1)$
&$(0.9\pm 0.1,\,0.7\pm 0.1)$\\
\hline
$\Xi_c^+ \to \Delta^{++} \pi^-$
&$-\lambda_cE_M$
&$(8.0\pm 1.5,\,7.8\pm 1.6)$
&$(25.5\pm 4.4,\,7.8\pm 0.7)$\\
$\Xi_c^+ \to \Delta^+ \pi^0$
&$-\lambda_c\frac{1}{\sqrt 6}(E_{\bf B}-E_M)$
&$(6.5\pm 1.0,\,6.3\pm 1.2)$
&$(17.0\pm 2.9,\,5.2\pm 0.4)$\\
$\Xi_c^+ \to \Delta^0 \pi^+$
&$-\lambda_c\frac{1}{\sqrt 3}E_{\bf B} $
&$(3.9\pm 0.6,\,3.7\pm 0.7)$
&$(8.5\pm 1.5,\,2.6\pm 0.2)$\\
$\Xi_c^+ \to \Delta^+ \eta$
&$-\lambda_c\frac{1}{\sqrt 6}(E_{\bf B}+E_M)c\phi$
&$(0.03\pm 0.32,\,0.03\pm 0.52)$
&-----\\
$\Xi_c^+ \to \Delta^+ \eta^\prime$
&$-\lambda_c\frac{1}{\sqrt 6}(E_{\bf B}+E_M)s\phi$
&$(0.01\pm 0.15,\,0.01\pm 0.24)$
&-----\\
$\Xi_c^+ \to \Sigma^{*+} K^0$
&$-\lambda_c\frac{1}{\sqrt 3}E_M^{(s)}$
&$(2.1\pm 0.4,\,4.2\pm 0.8)$
&$(3.5\pm 0.6,\,2.6\pm 0.2)$\\
$\Xi_c^+ \to \Sigma^{*0} K^+$
&$-\lambda_c\frac{1}{\sqrt 6}E_{\bf B}^{(s)}$
&$(1.5\pm 0.2,\,1.5\pm 0.2)$
&$(1.7\pm 0.3,\,1.3\pm 0.1)$\\
\hline
\end{tabular}
\end{table}

Uniquely, the decuplet baryon can contain three identical quarks, denoted by ${\bf B}^*(qqq)$,
which leads to an additional weight factor of $\sqrt 3$
among the decuplet baryons in Eq.~(\ref{B_10}). The factor can be considered
as the main reason why $\Lambda_c^+ \to \Delta^{++} K^-$ and
$\Xi_c^0 \to \Omega^- K^+$ are measured with the largest branching fractions
in the CA decay channels of $\Lambda_c^+,\Xi_c^0\to{\bf B}^*M$, respectively.
Accordingly, the $T$-amps with ${\bf B}^*(qqq)$ are listed as
\begin{eqnarray}
&&T(\Lambda_c^+\to \Delta^{++} K^-,\Delta^{++} \pi^-)=-(\lambda_a,\lambda_d)\,E_M\,,\nonumber\\
&&T(\Xi_c^+ \to \Delta^{++} K^-,\Delta^{++} \pi^-)=-(\lambda_s,\lambda_c)\,E_M\,,\nonumber\\
&&T(\Xi_c^0 \to \Omega^- K^+,\Delta^- \pi^+)=(\lambda_a,-\lambda_c)\,E_{\bf B}\,.
\end{eqnarray}
While ${\cal B}(\Lambda_c^+\to \Delta^{++} K^-)$
and ${\cal B}(\Xi_c^0 \to \Omega^- K^+)$ have been observed,
the other branching fractions are given by
\begin{eqnarray}
&&{\cal B}(\Lambda_c^+\to \Delta^{++} \pi^-,\Xi_c^+ \to \Delta^{++} K^-)
=(7.0\pm 1.4,13.5\pm 2.7)\times 10^{-4}\,,\nonumber\\
&&{\cal B}(\Xi_c^+ \to\Delta^{++} \pi^-,\Xi_c^0 \to \Delta^- \pi^+)
=(7.8\pm 1.6,2.5\pm 0.4)\times 10^{-5}\,,
\end{eqnarray}
which are predicted as the largest branching fractions in the SCS $\Lambda_c^+(\Xi_c^+)$ and
DCS $\Xi_c^{+(0)}$ decay channels, respectively. Here,
we present our predictions of $S2$, which is favored by the $\chi^2$-fit.
The equality of
$T(\Lambda_c^+ \to \Sigma^{*0} \pi^+)=T(\Lambda_c^+ \to \Sigma^{*+} \pi^0)$
corresponds to the isospin symmetry. 
The branching fraction, given by
\begin{eqnarray}
{\cal B}(\Lambda_c^+ \to \Sigma^{*0(+)} \pi^{+(0)})&=&(2.8\pm 0.4)\times 10^{-3}\,,
\end{eqnarray}
can be used to test the broken effect. The decays ${\bf B}_c\to {\bf B}^*\eta^{(\prime)}$,
$\Lambda_c^+ \to \Sigma^{*+} \eta$, $\Xi_c^0 \to \Xi^{*0} \eta$
and $\Xi_c^+ \to \Sigma^{*+}\eta$ have sizeable branching fractions,
which is due to the constructive interferences between $E_{\bf B}$ and $E_M$.
However, the other branching fractions of ${\bf B}_c\to {\bf B}^*\eta^{(\prime)}$
are typically small with the destructive interferences. Moreover,
we find that ${\cal B}(\Lambda_c^+ \to \Sigma^{*+} \eta^\prime)=0$
with $m_{\Lambda_c^+}<m_{\Sigma^{*+}}+m_{\eta^\prime}$.

The approach of the irreducible $SU(3)_f$ symmetry has been widely used in
the hadron weak decays~\cite{Savage:1989qr,Savage:1991wu,
He:2000ys,Fu:2003fy,Hsiao:2015iiu,Lu:2016ogy,Geng:2019awr,
Geng:2017esc,Geng:2018plk,Geng:2017mxn,Geng:2018upx,Hsiao:2019yur,Pan:2020qqo}.
For ${\bf B}_c\to{\bf B}^*M$,
there exist four parameters $a_8$ and $a_{9,10,11}$~\cite{Savage:1989qr,Geng:2017mxn},
which correspond to the decomposition of
${\cal H}_{eff}=H(6)+H(\overline{15})$
in the $SU(3)_f$ representation of $6$ and $\overline{15}$,
respectively. By comparison, we derive that
\begin{eqnarray}\label{ai}
(E_{{\bf B}},E_{M})=(-2 a_8+a_9,2a_8+a_9)\,,\;(E',C')=(-2a_9-2a_{10},-2a_{11})\,,
\end{eqnarray}
such that  $a_i$ are found to correspond to the topologies.
Since $(E',C')$ have been the vanishing topological parameters, 
one has $a_9=-a_{10}$ and $a_{11}=0$.
Moreover,
our global fits for $E_{{\bf B},M}$ indicate that $a_{9(10)}$ from $H(\overline{15})$
has a non-zero value.
By contrast, the numerical analysis performed with the irreducible $SU(3)_f$ symmetry
neglects the contributions from $H(\overline{15})$~\cite{Geng:2019awr},
whose results are given in the tables. In the physical mass scenario ($S_{pm}$)
for the global fit in Ref.~\cite{Geng:2019awr}, where
$m_{{\bf B}_c}$, $m_{{\bf B}^*}$ and $m_M$ are taken from the physical values in Ref.~\cite{pdg},
${\cal B}(\Lambda_c^+ \to \Sigma^{*+} \eta,\Xi^{*0} K^+)$
and ${\cal B}(\Xi_c^0 \to \Omega^- K^+)$
are fitted to be a few times smaller than the observations. 
Instead of considering the $SU(3)_f$ symmetry breaking,
one performs another fit in the equal mass scenario ($S_{em}$),
where $m_{\Lambda_c}=m_{\Xi_c}$, $m_\Delta=m_{\Sigma^*}=m_{\Omega}$
and $m_\pi=m_\eta=m_K$,
resulting in the raised values of the above branching fractions.

\section{Conclusions}
In summary, we have studied the ${\bf B}_c\to {\bf B}^* M$ decays
in the quark-diagram scheme. We have found that
only two $W$-exchange diagrams, $E_{\bf B}$ and $E_M$,
could give contributions to the observed branching fractions of
$\Lambda_c^+ \to (\Delta^{++}  K^-$, $\Sigma^{*+} \eta$, $\Xi^{*0} K^+)$ and
$\Xi_c^0 \to \Omega^- K^+$.
In addition, we have predicted
${\cal B}(\Lambda_c^+ \to \Sigma^{*0(+)} \pi^{+(0)})=(2.8\pm 0.4)\times 10^{-3}$,
which respects the isospin symmetry.
To interpret the observation of ${\cal B}(\Lambda_c^+ \to \Sigma^{*+} \eta)$,
we have taken into account the $SU(3)_f$ symmetry breaking.
In particular,
${\cal B}(\Lambda_c^+\to \Delta^{++} \pi^-,\Xi_c^+ \to \Delta^{++} K^-)
=(7.0\pm 1.4,13.5\pm 2.7)\times 10^{-4}$ and
${\cal B}(\Xi_c^+ \to\Delta^{++} \pi^-,\Xi_c^0 \to \Delta^- \pi^+)
=(7.8\pm 1.6,2.5\pm 0.4)\times 10^{-5}$ have been predicted as
the largest branching fractions in the SCS $\Lambda_c^+(\Xi_c^+)$ and
DSC $\Xi_c^{+(0)}$ decay channels, respectively.

\newpage
\section*{ACKNOWLEDGMENTS}
We would like to thank Prof. X.G. He for useful discussions.
This work was supported 
by National Science Foundation of China (11675030).

\end{document}